# Reusability Framework for Cloud Computing

## Sukhpal Singh[1], Rishideep Singh[2]

[1] M.E. (S.E.) Computer Science and Engineering Department, Thapar University, Patiala, India,
[2] Assistant Professor, Department of Information Technology, N.W.I.E.T. Dhudike, Moga, India.

**Abstract:**
Cloud based development is a challenging task for several software engineering projects, especially for those which needs development with reusability. Present time of cloud computing is allowing new professional models for using the software development. The expected upcoming trend of computing is assumed to be this cloud computing because of speed of application deployment, shorter time to market, and lower cost of operation. Until Cloud Computing Reusability Model is considered a fundamental capability, the speed of developing services is very slow. This paper spreads cloud computing with component based development named Cloud Computing Reusability Model (CCR) and enable reusability in cloud computing. In this paper Cloud Computing Reusability Model has been proposed. The model has been validated by Cloudsim and experimental result shows that reusability based cloud computing approach is effective in minimizing cost and time to market.

***Keywords:*** Cloud based software development, Component based development, Cloud Component, Cloud computing, Reusability, Software engineering, Software reuse.

## 1. Introduction

Reusability means using a segment of source code that can be used again to add new functionalities with slight or no modification. In most engineering disciplines, systems are designed by composing existing components that have been used in other systems [26]. Software engineering has been more focused on original development but it is now recognized that to achieve better software, more quickly and at lower cost, we need to adopt a design process that is based on systematic software reuse [1]. Reverse engineering means evaluating something to understand how it works in order to duplicate or enhance it. It allows the reuse of the know-how hidden inside already implemented programs [12] [14]. The object oriented software developers now admit that thinking about object-oriented program understanding and comprehension to be relatively easier is not that easy. Programs are even more complex and difficult to comprehend, unless rigorously documented. What if the documentation is improper? To affect change management, even a simpler upgrade may become cumbersome then [3] [25]. This is the reason why eminent development houses now focusing on advanced documentation support [39]. Re-engineering code environment hence largely affect the problem issues regarding program comprehension when the software size grows enormously. Reverse Engineering is a methodology that greatly reduces the time, effort and complexity in solving these issues providing efficient program understanding as an integral constituent of re-engineering paradigm [2] [26]. Cloud computing is the use of computing resources (hardware and software) that are delivered as a service over a network (typically the Internet) [33] [38]. The name comes from the use of a cloud-shaped symbol as an abstraction for the complex infrastructure it contains in system diagrams. Cloud computing [11] [30] entrusts remote services with a user's data, software and computation [13] shown in Figure 1. In Section 2 the related work has been described. The challenges of cloud computing platform for software is analyzed in Section 3. In Section 4 the Cloud Computing Reusability Model (CCR) has been discussed. The experimental results are explained in Section 5. The advantages of proposed model have been discussed in Section 6. The Section 7 concludes the whole work and provides future work in Section 8.

## 2. Related work

Reverse engineering is a systematic form of program understanding that takes a program and constructs a high-level representation useful for documentation, maintenance, or reuse. To accomplish this, reverse engineering technique begins by analyzing a program's structure [24]. The structure is determined by lexical, syntactic, and semantic rules for legal program construction. Because we know how to proceed on these kinds of analysis, it is natural to try and apply them to understand programs. Initially reverse engineering term was evolved in the context of legacy software support but now has ventured into the important issue of code security such that it doesn't remain confined to legacy systems. We will come to the discussion into this effect after a while. Transformations are applied under the process of Re-engineering [25] after analyzing the software to apply changes incorporating new features and provide support for latest environment. Object-oriented software development methodology primarily has three phases of Analysis, Design and Implementation [36]. With the view of the traditional waterfall model, reverse engineering this is looking back to design from implementation and to analysis from implementation. The important thing is that it actually is a reverse forward engineering i.e. from implementation; analysis is not reached before design. The software system or program under study is neither modified nor re-implemented because of not bringing it under Re-engineering. [23] Software Re-engineering is the area which deals with modifying software to efficiently adapt new changes that can be incorporated within as software aging is a well-known issue. Reverse engineering provided cost effective solutions for modifying software or programs to adapt change management through Re-engineering application [25] [10]. Reusable architectures [31] can be developed from reusable architectural patterns [11] as in FIM architecture [12] which operates at three different levels of reuse: Federation, domain and application. Paulisch F. et. al. focuses on how non-functional property reusability relates to the software architecture (SOA) of a system [4] [7]p. K.S. J. and Dr. Vasantha R. presented a software process model for reuse based software development approach [5] [17] [18]. From the overall literature survey, it can be concluded that: code, design, test cases etc can be reused. Reuse can be systematic (software development for reuse), or opportunistic (software development with reuse) Reuse does not just happen; it needs to be planned and require proper documentation and design. Reverse-engineering can be used for reusability or it can be said that reusability can be achieved using reverse-engineering. Reverse engineering helps to understand the legacy system by creating its UML model and once the model of the legacy system is created, that model can be used with little or no modification in the underdevelopment or in the future project to promote reusability and to increase productivity of the organization [6] [12]. There are lots of UML tools available to perform reverse-engineering process [9] [15]. Reverse-Engineering can be used to make the poorly designed and poorly documented legacy software system developed with cloud development process; Re-usable by extracting the component from the legacy system using UML models [8] [26].

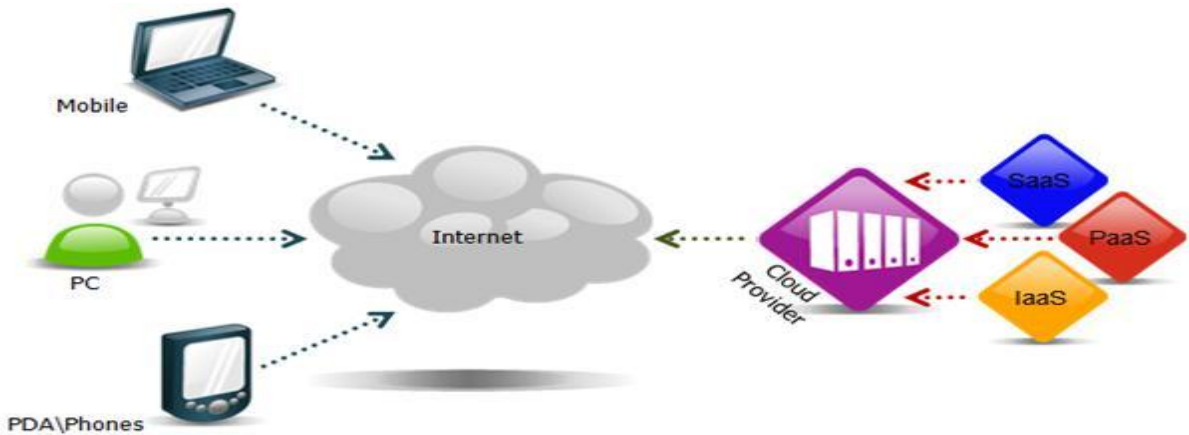

**Figure 1. Cloud Computing and its services [13]**

Reuse based software engineering and cloud development is an open research area in rapid development. We had conducted a survey on the number of approaches existing for Cloud Based Development [14, 9], and Reusability [11] [16] individually, but the proposed model combines both Cloud computing and Reusability [19] into a single approach for achieving efficient classification, storage and retrieval of software components and improve time to market and reduce cost. The cost without reusability is increasing phase to phase as shown in Figure 2. Presently there is no such approach as presented in proposed model which combines the Component based Development (Reusability) [24] [29] and Cloud computing [15].

## 3. Analysis

In the rapidly changing computing environment with cloud platform, software development is going to be very challenging. The software development process will involve heterogeneous platforms, distributed web services, multiple enterprises geographically dispersed all over the world. Figure 2 shows the cost of development is increasing from requirement to maintenance without reusability for a small project, the development with reuse will cut an initial cost and reduce time to market. The organizations need to incorporate reuse in their development [20] [32]; it will be a long term investment process. Figure 3 summarizes incremental reuse levels (solid line) and related reuse approaches (dotted lines) [21]. The reuse of ad-hoc reuse events with initial benefits is the only level of reuse which can be achieved without investing in software reuse; instead the experience of previous projects is used to copy relevant pieces of code.

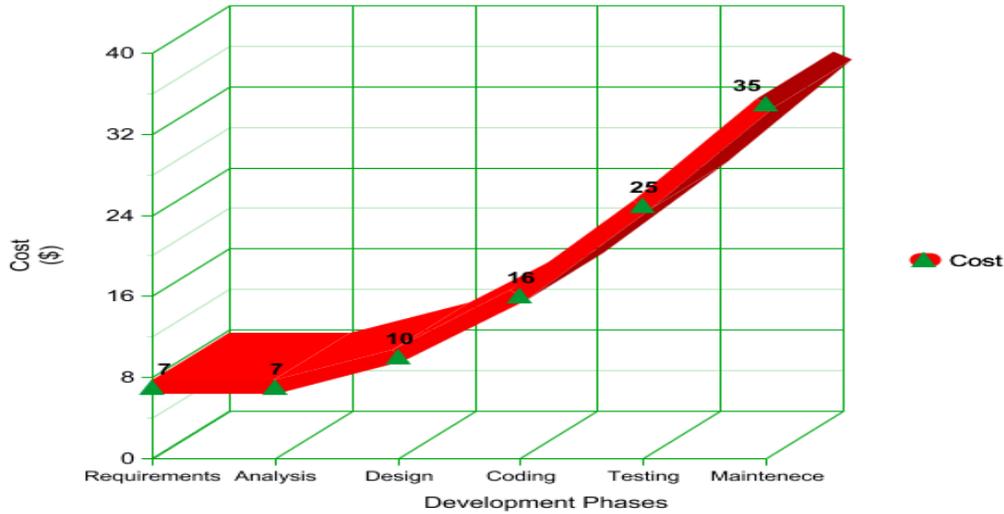

**Figure 2. Evolution of Software Engineering**

This reuse level is defined as level 0. The first real reuse level presents a form of code-leverage, where pieces of code are made available and can be reused by multiple parties. The pieces of code are made available through the use of a reuse library, providing a central place where the components are stored [37].

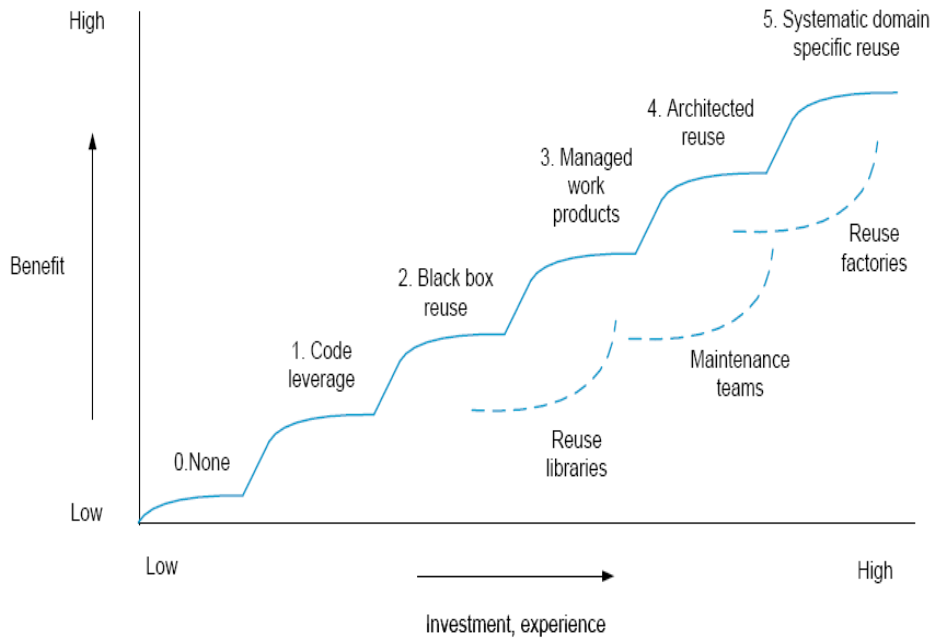

**Figure 3. Incremental stages of reuse [21]**

The use of these leveraged components is not restricted and can be considered as white-box reuse; the code may be adjusted and changed to the specific context in which the component is applied [27] [35]. This strategy works for a while up to the point that multiple copies, each slightly different, have to be managed. The choice can be made to stop using components as white-box and start using them as black-box components instead. Black-box components may no longer be internally adjusted; rather its environment should be adjusted to support the component. Previous research has pointed out that with black-box reuse higher reuse levels can be achieved than with white-box reuse [22] [36]. The reason for this is that there are reduced costs in component maintenance and maintenance across products using these components. However, black-box reuse also has its limitations.

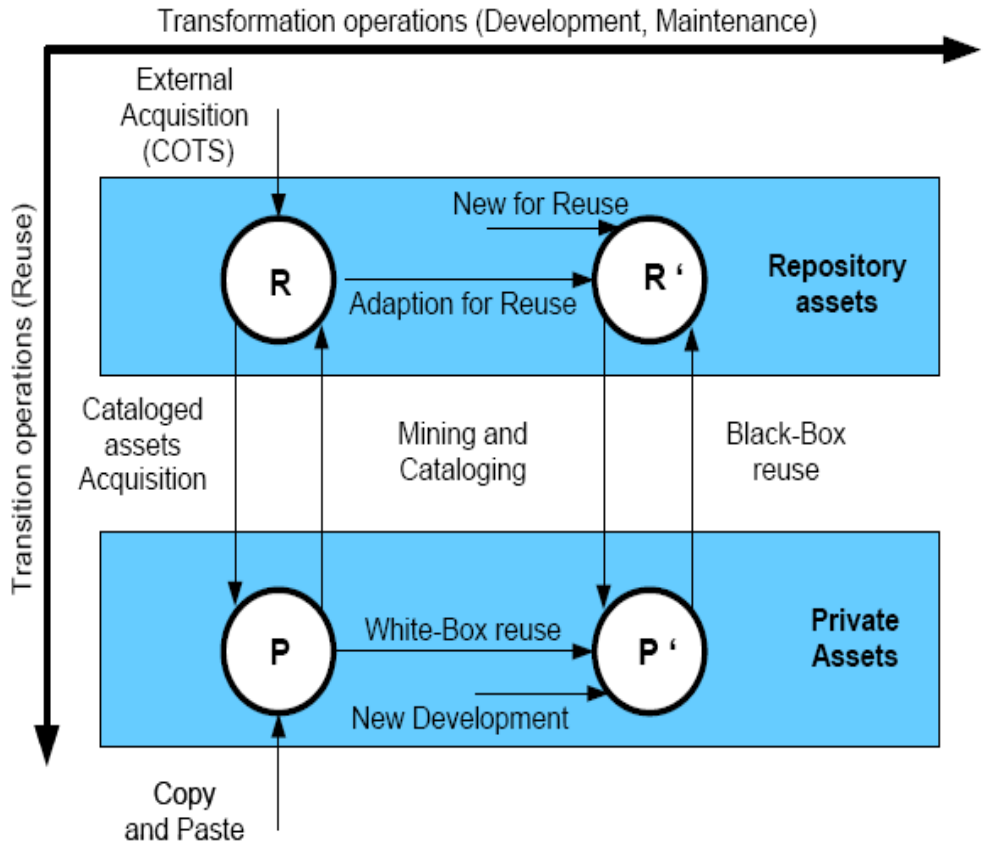

**Figure 4. Assets and reuse operations [22].**

The axis of the three dimensional model are explained by an underlying model, also presented in Tomer et al. [22]. The underlying model is presented in Figure 4. For simplicity Tomer et al. transferred the underlying model to a two dimensional model, where development and maintenance are combined into one axis [34]. A key assumption made in this model is that reuse activities cannot be freely transferred between specific products without first storing and cataloguing the assets in a central repository [22]. Hence incorporating reuse in cloud computing will be beneficial if the general component will be used multiple times for developing cloud projects and services.

**4. Cloud Computing Reusability (CCR) Model**

Innovative software engineering is required to leverage all the benefits of cloud computing and mitigate its challenges strategically to push forward its advances. Here we propose an extended version of software development, reusability process model for cloud computing platform and name it Cloud Computing Reusability (CCR) Model [Figure 5]. A model capable of developing cloud based applications with reusability by retrieving the components from the cloud component repository by using pattern matching algorithms and various retrieval methods. There are some different 6 retrieval methods available for the classification of components in the software library. This model will help to make searching faster based on classification of components and develop the cloud based application according to cloud customer requirements to improve time to market as compared to traditional software development approach. The Cloud Component Architecture of proposed Cloud Computing Reusability (CCR) Model is shown in Figure 6. In the Unified Modeling Language, a component diagram depicts how cloud components are wired together to form larger component. They are used to illustrate the structure of arbitrarily complex cloud systems.

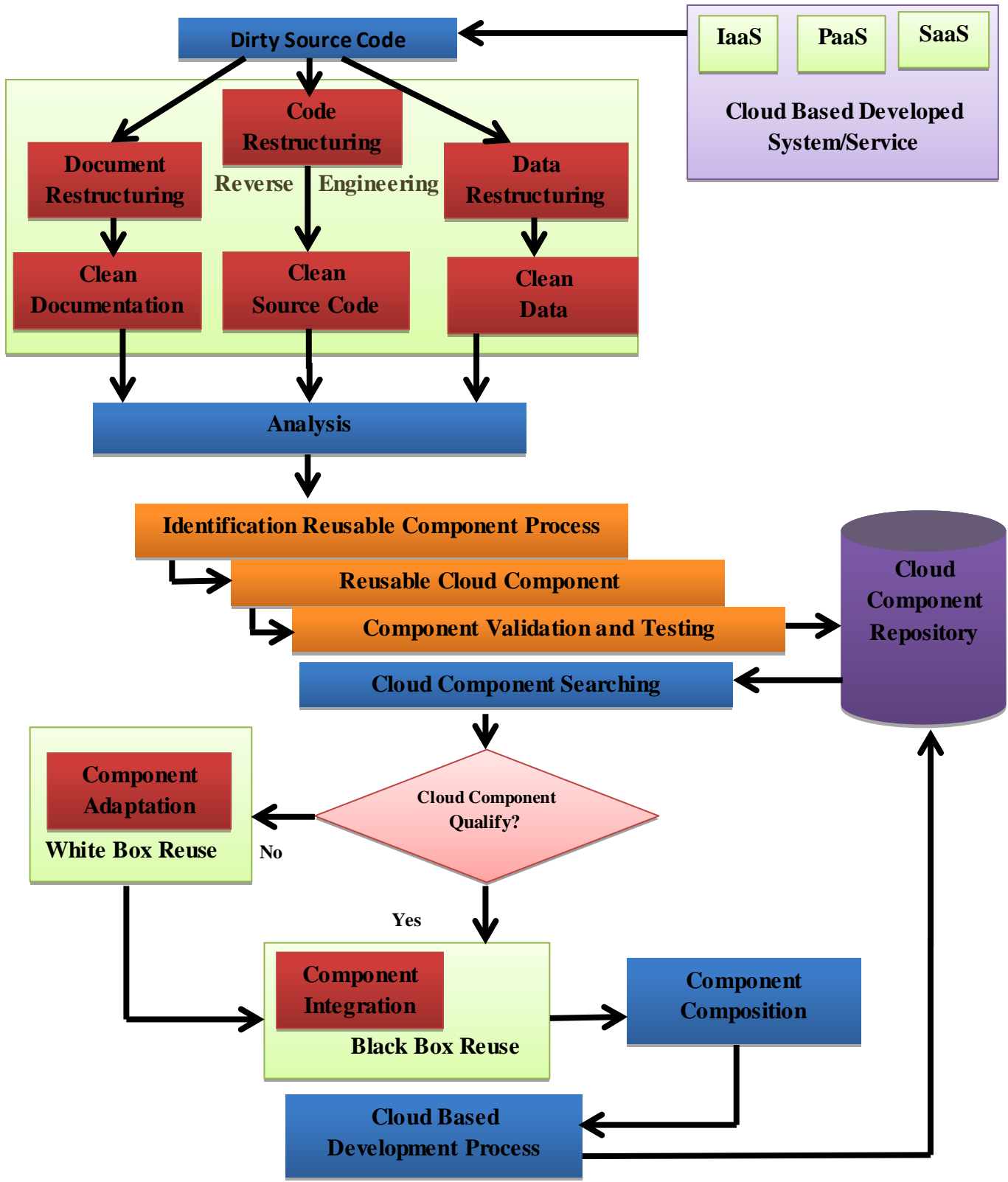

**Figure 5. Cloud Computing Reusability (CCR) Model**

### 4.1 Cloud Based Developed System/Service

In cloud computing, three services are provided to the customer; Infrastructure as a Service, Platform as a Service and Software as a Service. The organization reuses their developed projects by black box reuse and if the cloud service or project is developed by another organization then it will be used by reverse engineering in cloud computing then if it will be further updated by white box reuse.

### 4.2 Reverse Engineering

The dirty source code is obtained from cloud based development and service and then code, data and document restructuring will be performed to find a clean document (UML), data (Meta Data) and Code (coding style). After that it will be analyzed for future use, to check whether it is feasible or not, if feasible then identify reusable process. The reusable cloud component will be obtained through this process and then validation and verification of cloud component will be performed and then sent to the cloud component repository.

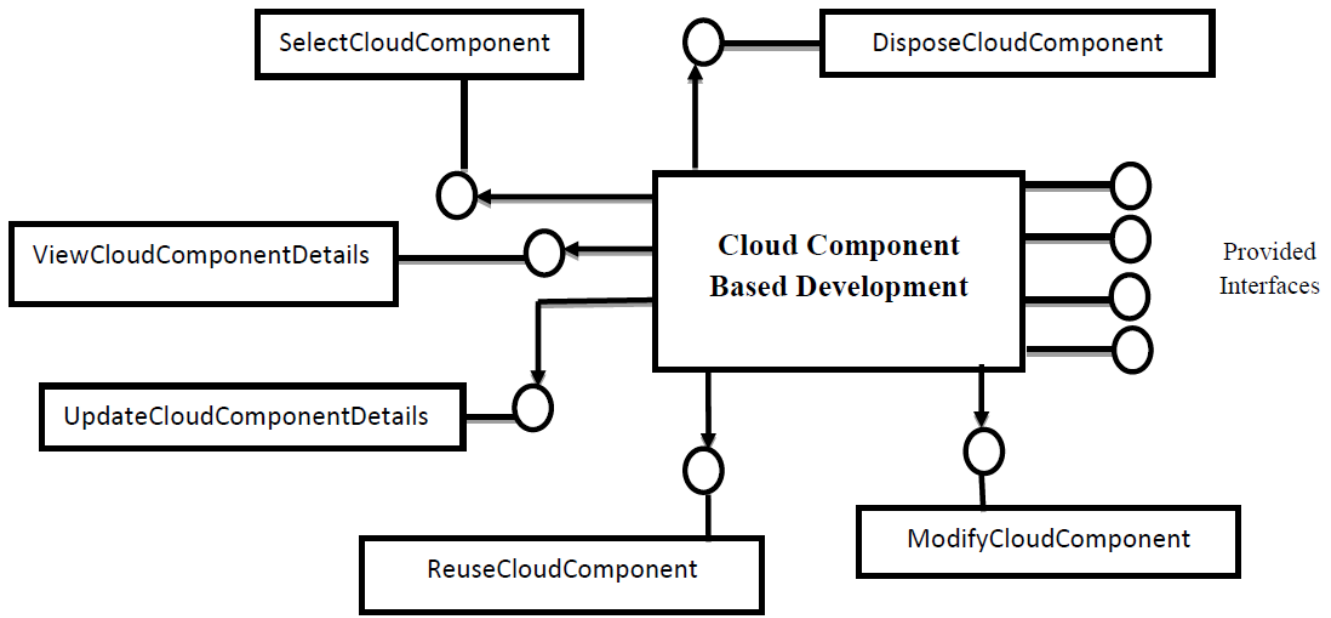

**Figure 6. Cloud Component Architecture**

### 4.3 Cloud Component Repository

The refined cloud developed components by reverse engineering with traditional development model will be stored in cloud component repository and retrieved it at a later stage for other cloud application development. There is some different storage and retrieval methods (Information retrieval methods, Operational semantics methods, Descriptive methods, Denotational semantics methods, Topological methods and Structural methods) are available for the classification of components in the software library. This model will help to make searching faster based on classification of cloud components and use these cloud components in other projects by searching and send back to cloud component repository after updating.

### 4.3 Cloud Component Reuse Process

The searched cloud component will be send through the phase cloud component qualification to check whether the component support required architecture, functionality and interfaces. If it qualifies then it will reused as a black box otherwise it will be reused as a white box reuse through the modification then the component will be integrated with current cloud application and send back to cloud component repository for future use.

### 5. Results and Discussion

The results of Cloud Computing Reusability (CCR) Model as compared to traditional cloud based development have been described in the Table 1. The result with Cloud Computing Reusability (CCR) Model has been verified with the help of Cloudsim.

**Table 1. Comparison of CCR method and traditional method**

| Criteria | CCR Method | Traditional Methods |
|---|---|---|
| Approach | Adaptive | Predictive |
| Success Measurement | Business Value | Confirmation To Plan |
| Project Size | Small | Large |
| Management Style | Decentralized | Autocratic |
| Perspective To Change | Change Adaptability | Change Sustainability |
| Culture | Leadership Collaboration | Command Control |
| Documentation | Low | Heavy |
| Emphasis | People Oriented | Process Oriented |
| Cycles | Numerous | Limited |
| Domain | Unpredictable/Exploratory | Predictable |
| Upfront Panning | Minimal | Comprehensive |
| Return On Investment | Early In Project | End Of Project |
| Team Size | Small/Creative | Large |
| Cost | Less | More |
| Time To Market | Less | More |

Figure 7 shows the HR application developments in cloud computing as a development for reuse, and then it will be four times used in four projects. The HR application which has been stored in cloud component repository, now used in cloud computing projects: MRI Imagining, Clinical Trial, Viral Marketing and Molecule Research.

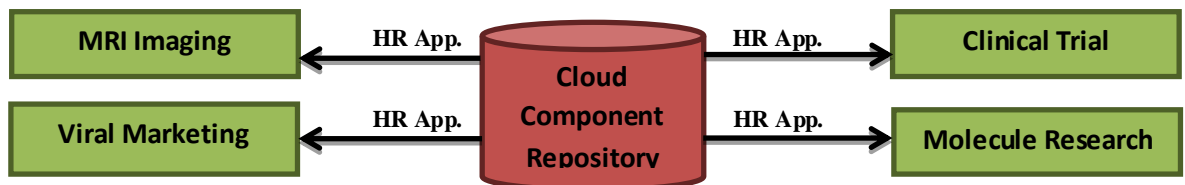

**Figure 7. Reuse of HR App. Cloud component in other projects**

The services provided by cloud provider will be easily reusable for new cloud projects. The evaluation of Cloud Computing Reusability (CCR) Model has been summarized in Table 1. The cost and time to market reduces in Cloud Computing Reusability (CCR) Model as compared to traditional software development, better services will be delivered to the cloud user with large satisfaction level, and these results were verified by Cloudsim. The cost reduction and improvement in time to market proved by Cloudsim shown in Figure 8.

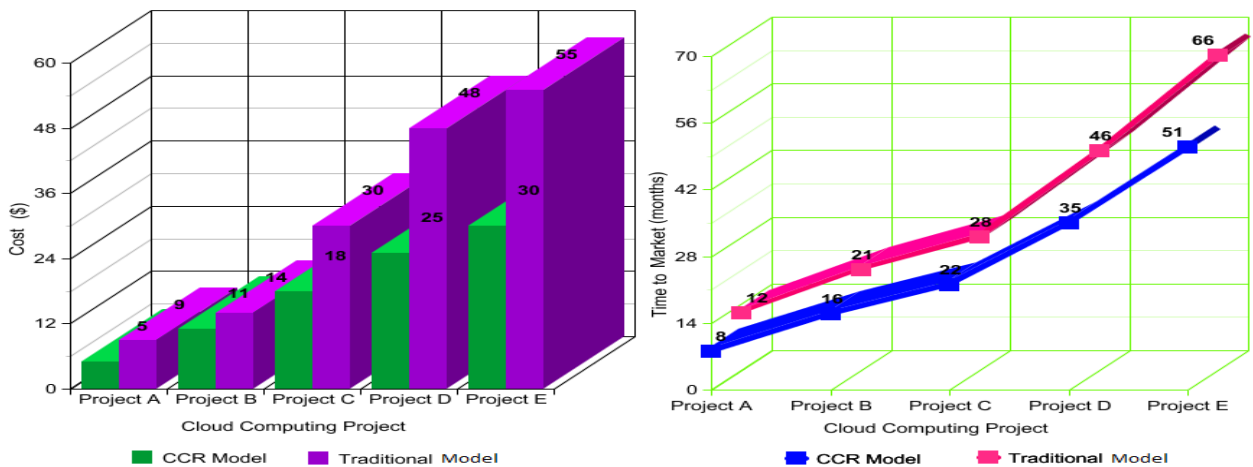

**Figure 8. The comparison of cost and time to market for Traditional and CCR Model**

## 6. Advantage of Proposed Approach

This proposed Cloud Computing Reusability (CCR) will help to 1) developing application quickly 2) reduces cost 3) improves reusability 4) reduce time to market and make searching faster based on classification of components and introducing reusability in Software Development. It will be accepted widely if pattern based architecture designing, design patterns [27] [28], UML based analysis and designing is incorporated. The six important ways Cloud Computing Reusability (CCR) enhances cloud based development.

- Enhance the productivity and improve the quality and reliability of the new software systems.
- Identify independent components having a low coupling and high cohesion.
- Accelerate cloud based development
- Improve time to market
- Reduce cost
- Increase generality

## 7. Conclusion

In this paper, we have presented a Cloud Computing Reusability (CCR) Model. The objective is to minimize the complexity, cost, time to market and increase reusability, development speed and generality. Software Development with reusability has encouraging future in the software industry and is capable of fulfilling the requirements of the cloud industry. Thus, at times it compromises with quality and is incapable of providing reusability of its cloud based developed components. Traditional Software Development offers particular solutions whereas Reuse and Cloud component based Development believe in generalized solutions to satisfy the demands of cloud customer. Component based development is a standard shift over the traditional way of developing and deploying of software. The amount of effort required for evolving software with reusability will diminish but there will be added communication and coordination requirement with the developer which makes software development project more difficult. The main objective of this paper is that the leading software process models should incorporate this new dimension of interaction with the reusability. A new Cloud Computing Reusability (CCR) Model is proposed in this paper which includes the expected communication requirement with the application developer and component developer which will diminish all the challenges of software development on a cloud computing platform and make it more beneficial to develop and deploy software on the cloud computing platform. The model is based on Reverse Engineering for identifying and creating reusable software component and reused that component. A model based on pattern matching technique is used to search the cloud component from the cloud component repository. This model encompasses the reverse engineering methodology to extract components of the object oriented legacy cloud system development. It utilizes cloud component repository to store and manage the tested components and restructures the new system that finally integrates the new system with the reusable components. The reusability of the cloud component is the most popular way to enhance the productivity and improve the quality and reliability of the new software systems by reducing the development costs. Due to these reasons, it is very important to identify independent components having a low coupling and high cohesion. Also a systematic approach to identify reusable component from the object oriented legacy system through cloud component architecture has been proposed. The proposed approach has been validated by using a UML and also its components are tested for reusability and illustrated that how these components can be reused in other cloud projects.

## 8. Future Work

The future scope of this work is to analyze and to incorporate risk factors in Component Based Development systematically and find the critical success factors of the Cloud Computing Reusability (CCR) and also identify the various risk factors using risk analysis of introducing reusability in component based development and offer a model that will help us to achieve reusability in Cloud Development. Reusability can also be automated in cloud development using an automated tool. Current results have been gathered through the simulation on Cloudsim but in future the same results would be verified actually by cloud providers. In future, if this proposed methodology can be fully automated by an automatic tool then it could be more effective and less time consuming. Component based software engineering and cloud computing is an open research area in fast growth.